\documentclass[aps,prl,twocolumn,amsmath,amssymb,superscriptaddress,reprint, floatfix]{revtex4}
\usepackage{times}
\usepackage{graphicx}
\usepackage{color}

\begin{document}

\title{A detailed analysis of the Raman spectra in superconducting boron doped nanocrystalline diamond}


\author{P\'{e}ter~Szirmai}
\affiliation{Faculty of Physics, University of Vienna, Strudlhofgasse 4., Vienna, A-1090, Austria}
\affiliation{Department of Physics, Budapest University of Technology and Economics, PoBox 91, Budapest, H-1521, Hungary}

\author{Thomas~Pichler}
\affiliation{Faculty of Physics, University of Vienna, Strudlhofgasse 4., Vienna, A-1090, Austria}

\author{Oliver~A.~Williams}
\affiliation{School of Physics and Astronomy, Cardiff University, Cardiff CF24 3AA, UK}

\author{Soumen~Mandal}
\affiliation{Institut N\'{e}el - CNRS and Universit\'{e} Joseph Fourier, 38042 Grenoble, France}

\author{Christopher~B\"{a}uerle}
\affiliation{Institut N\'{e}el - CNRS and Universit\'{e} Joseph Fourier, 38042 Grenoble, France}

\author{Ferenc~Simon}
\email[Corresponding author: ]{ferenc.simon@univie.ac.at}
\affiliation{Department of Physics, Budapest University of Technology and Economics, PoBox 91, Budapest, H-1521, Hungary}






\date{\today}
\begin{abstract}
The light scattering properties of superconducting ($T_{\text{c}}\approx3.8$ K) heavily boron doped nanocrystalline diamond has been investigated by Raman spectroscopy using visible excitations. Fano type interference of the zone-center phonon line and the electronic continuum was identified. Lineshape analysis reveals Fano lineshapes with a significant asymmetry ($q\approx -2$). An anomalous wavelength dependence and small value of the Raman scattering amplitude is observed in agreement with previous studies.
\end{abstract}

\maketitle
\section{Introduction}
The diversity of carbon materials gives rise to remarkable range of applications. Probing these materials require a non-destructive method with little preparation. Raman spectroscopy is a powerful diagnostic tool to study nanocarbons ranging from graphene \cite{FerrariPRL2006}, single-walled carbon nanotubes \cite{DresselhausCNTRamanReview}, to diamond-like materials \cite{BDD_Prawer}.\\
Diamond possesses a number of unique properties (such as e.g.\ the well known hardness, large tensile strength, and thermal conductivity) which may lead to a unique class of diamond based integrated devices \cite{BDD_Mandal}. Due to its high electrical conductivity and chemical stability, boron doped diamond (BDD) has been widely studied \cite{BDD_Field}. The discovery of superconductivity in heavily boron doped diamond \cite{BDD_Nature2004} led to renewed interest. Apart from crystals synthesized at high pressure and high temperature \cite{BDD_Nature2004}, bulk superconductivity was soon found in diamond films grown with the microwave plasma-assisted chemical vapor deposition method (MPCVD) \cite{BDD_Bustarret2004}. In MPCVD samples, $T_{\text{c}}=11$ K as high as was measured \cite{BDD_CVD_Diamond2007}, however, this value is far from the highest prediction of $T_{\text{c}}\approx 55$ K \cite{BDD_Moussa}.\\
BDD was confirmed to be an example of Mott's metal \cite{Mott1968} above the threshold carrier concentration of $n_c\approx 2-3\cdot 10^{20}$ $\text{cm}^{-3}$ \cite{BDD_Gonon,BDD_KleinPRB,BDD_Bustarret2008}. The onset of metallic conductivity (and the superconductivity) in BDD can be followed by the change of the zone-center phonon (ZCP) Raman peak of diamond \cite{BDD_Ushizawa,BDD_Pruvost}. Interference between the ZCP and the continuum of electronic transitions can occur in the Raman signal of BDD \cite{BDD_Gheeraert1993}. This is known as the Fano effect \cite{Fano1961} and it was also found in heavily doped silicon \cite{BDD_Cerdeira}.\\
The position of the Fano resonance peak is widely used for the calibration of the boron content \cite{BDD_Nature2004}. Nevertheless, the origin of the Fano line is still controversial in BDD \cite{BDD_Pruvost,BDD_Ager,BDD_Wang2002}. In order to gain a better understanding of this Raman mode, a detailed characterization is needed. In Ref.~\cite{BDD_Pruvost}, an unusual wavelength dependence of the Raman scattering amplitude of the decoupled phonon ($T_{\text{p}}$) was found. In addition, the asymmetry of the lineshape is affected by the energy of the excitation \cite{BDD_Locher,BDD_Ghodbane}.\\
Herein, we report on Raman measurements in BDD using visible excitations. The origin of Raman modes in nanocrystalline BDD is assessed. Based on a Fano lineshape analysis, the anomalous behaviour of $T_{\text{p}}$ is confirmed in the visible range. Additionally, a smaller upper bound is found for $T_{\text{p}}$ than that given in Ref.~\cite{BDD_Pruvost}.

\section{Experimental}
Silicon $(111)$ wafers were cleaned by standard RCA SC1 processes. Diamond nucleation was initiated by immersion of clean wafers in aqueous colloids of hydrogenated nanodiamond particles in an ultrasonic bath. This process is known to produce nucleation densities in excess of 10$^{11}$ cm$^{-2}$ \cite{BDD_Hees}. Diamond growth by Microwave Plasma Enhanced Chemical Vapour Deposition was performed with 4\% CH$_4$ diluted in H$_2$ with 6400~ppm of trimethylboron \cite{BDD_Gajewski}. The microwave power was 3 kW and the substrate temperature of 800 $^{\circ}$C as monitored by a Williamson Pro92 dual wavelength pyrometer. The growth duration was 20 hours, which yields films of approximately 6~$\mu$m thickness. Transport experiments found $T_{\text{c}}=3.8\,$ K.\\
Raman spectra were recorded on a modified broadband LabRAM spectrometer (Horiba Jobin-Yvon Inc.). The built-in interference filter was replaced by a broadband beam splitter plate with 30 $\%$ reflection and 70 $\%$ transmission. The principles of the broadband operation are described elsewhere \cite{FabianRSI,FabianPSSB}. The spectrometer was operated with a 600 grooves/mm grating. A typical 1~mW laser power was used with a built-in microscope (Olympus LMPlan 50x/0.50 inf./0/NN26.5) which yields about $1\times1\text{ }\mu\text{m}^2$ spot size.
\section{Results and discussion}
\begin{figure}[tb]
\includegraphics*[width=\linewidth]{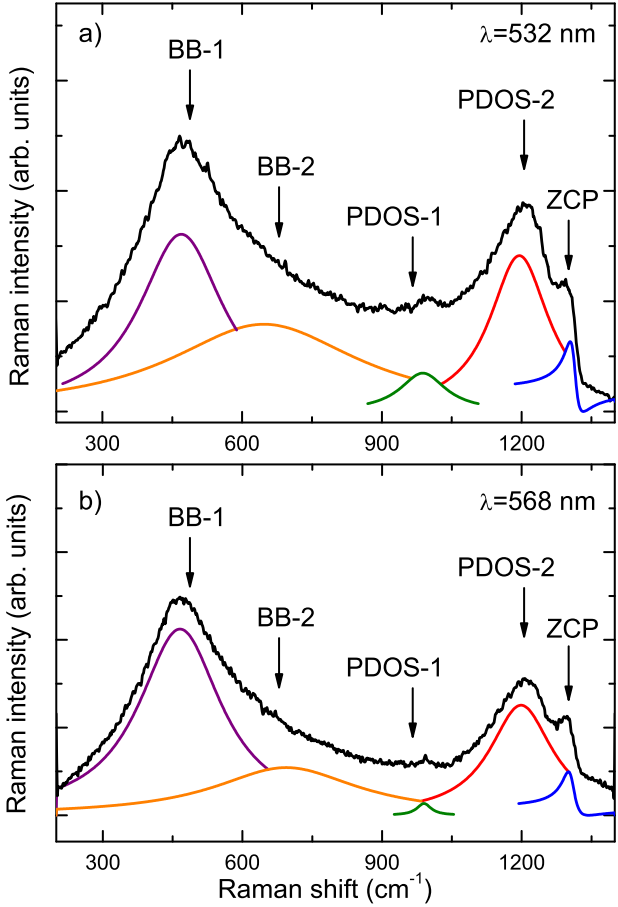}
\caption{Raman spectra of BDD  \textit{a)} at $\lambda=532~\text{nm}$ and \textit{b)} at $\lambda=568~\text{nm}$. Labels denote the boron dimers or point defect states (BB-1 and BB-2), the peaks due to the maxima of the phonon density of states (PDOS-1, PDOS-2), and the zone-center phonon line (ZCP). Solid Lorentzian (BB-1, BB-2, PDOS-1 and PDOS-2) and Fano (ZCP) lines show the fit for the spectra.}

\label{RamanFano}
\end{figure}
In Fig.~\ref{RamanFano}, Raman spectra of heavily boron doped diamond (BDD) are depicted at $\lambda=532~\text{nm}$ and $\lambda=568~\text{nm}$.\\
The Raman bands around $500$ cm$^{-1}$ (denoted by BB-1 and BB-2 in Fig.~\ref{RamanFano}), were assigned to boron dimers \cite{BDD_Bourgeois,BDD_Bernard2004_DandRelMat,BDD_Sidorov2010} and to clustered boron atoms \cite{BDD_Sidorov2010}. 
Using first-principles calculations, boron dimers were found to form a stable structure upon heavy doping \cite{BDD_Bourgeois}. The dimer-related vibrational $A_{1g}$ and $E_{g}$ modes occur in our spectra as a broad feature.\\
As the low energy of the phonon mode would reflect unusually weak force constant between boron atoms, an alternative assignment is used as well in the literature \cite{BDD_Prawer}. Diamond irradiated with high-energy alpha particles showed the role of damage and defect concentration on Raman spectra \cite{BDD_Orwa}. Amorphous diamond exhibits similar broad feature as heavily boron doped diamond.\\
Isotopic substitution did not give satisfying identification of the origin of the feature as both boron and carbon substitution gives shift on Raman spectra \cite{BDD_Sidorov2010}. The controversy of the interpretation might be resolved by regarding boron dimers to be point defects present in BDD in high concentration. The defect concentration correlates with the incorporated boron concentration, which reaches its maximum on the surface of the nanocrystals \cite{BDD_Liao}. The inhomogeneous distribution of boron atoms in BDD \cite{BDD_Li} leads to the breaking of the Raman wavevector conservation rule and yields similar Raman spectra for BDD as for amorphous diamond.\\
The peak was fitted with the sum of two Lorentzian components (fit shown in Fig.~\ref{RamanFano}), instead of a Lorentzian and a Gaussian component \cite{BDD_Bernard2004_DandRelMat}. The empirical relationship between the Raman shift of the lower Lorentzian component and the boron content ($n$) measured by secondary ion mass spectrometry (SIMS) found in Ref.~\cite{BDD_Bernard2004_DandRelMat} yields $n\approx 1.8\cdot 10^{21}$ $\text{cm}^{-3}$ for the boron content in our sample. This value is in agreement with that obtained with Hall effect measurements in a similar material \cite{BDD_Gajewski}. The approximate agreement found here and elsewhere \cite{BDD_May} further proves that the mode might be due to the high boron dimer concentration.\\
The Raman band around 1000 $\text{cm}^{-1}$ (PDOS-1)  originates from the maximum of the phonon density of states of diamond. As discussed above, defects in the material make the otherwise forbidden states allowed \cite{BDD_Prawer1998}. The Raman structure around 1200 cm$^{-1}$ consists of two components: a Lorentzian like at 1210 cm$^{-1}$ and another one with an asymmetric lineshape around 1300 cm$^{-1}$. The 1210 cm$^{-1}$ (PDOS-2) mode appears due to the presence of defects \cite{BDD_Sidorov2010,BDD_Vlasov}. In Ref.~\cite{BDD_Pruvost}, the mode is assigned to boron-carbon complexes. It was reported that this peak exhibits Fano lineshape \cite{BDD_Ghodbane}. However, the mode does not exhibit asymmetry in our Raman spectra.\\
The zone center optical phonon of diamond, which occurs at 1332 cm$^{-1}$ with $\gamma=1.2$ cm$^{-1}$ linewidth \cite{BDD_Liu}, is shifted to 1300-1313 cm$^{-1}$ in BDD and acquires a Fano lineshape due to the presence of free charge carriers \cite{BDD_Pruvost,BDD_Gheeraert1993}.\\
In Fig.~\ref{RamanFano}, we show Fano lineshapes fitted on Raman spectra. The Fano lineshape \cite{Fano1961} originates from the quantum interference between the zone center optical phonon and a continuum of electronic transitions around the same energy. It can be calculated as
\begin{equation}
\text{Int}(\omega)\propto \frac{\left[q+\left(\frac{\omega-\Omega_0}{\varGamma}\right)\right]^2}{1+\left(\frac{\omega-\Omega_0}{\varGamma}\right)^2},
\label{Fano_calc}
\end{equation}
where Int$(\omega)$ is the intensity of the Raman signal, $\hbar\Omega_0$ and $\hbar\varGamma$ are respectively the real and imaginary parts of the self-energy in BDD after coupling between discrete phonon transition and a continuum of states:
\begin{equation}
\hbar\varGamma=\pi V^2 \varrho(E)+\hbar\gamma,
\label{gamma_calc}
\end{equation}
and
\begin{equation}
\hbar\Omega_0=\hbar\omega_0-V^2R(E).
\label{omega_calc}
\end{equation}
Herein, $\varrho(E)$ is the density of states of the quasi-continuum, $\hbar\omega_0$ and $\hbar\gamma$ are respectively real and imaginary part of the self-energy of the phonon in the undoped material. $V$ is the matrix element of the coupling between the two states and $\pi^{-1}R(E)$ is the Hilbert transform of $V$.\\
The asymmetry parameter, $q$, reads
\begin{equation}
q=\frac{V\frac{T_{\text{p}}}{T_{\text{e}}}+V^2R(E)}{\pi V^2 \varrho(E)},
\label{q_calc}
\end{equation}
\begin{table}[!t]
  \caption[]{  \label{Fano_summary} Parameters of the Fano lineshapes. Data denoted by asterisks are from Ref.~\cite{BDD_Pruvost}.}
  \begin{tabular}{l|cccc}
  \hline\noalign{\smallskip}
   $\lambda$ (nm)&$q$&$\Omega_0$ (cm$^{-1}$)&$\Gamma$ (cm$^{-1}$)& $V\cdot T_{\text{p}}/T_{\text{e}}$ (meV)\\
   \hline\noalign{\smallskip}
   	532&-1.4&1313&14.1&-4.6\\
		568&-2.3&1307&15.7&-7.2\\
		514$^{*}$&-1.7&1320&15.6&-4.5\\
		633$^{*}$&-2.4&1316&14&-5.8\\
		\noalign{\smallskip}\hline
  \end{tabular}
\end{table}
where $T_{\text{p}}$ and $T_{\text{e}}$ are the Raman scattering amplitudes of the decoupled phonon and the electronic continuum \cite{BDD_Pruvost,BDD_Cerdeira}. If the electronic continuum disappears, and $q\rightarrow\infty$, the Fano formula becomes the usual Lorentzian curve.\\
Parameters of the Fano resonance lines are given in Table~\ref{Fano_summary}. Therein, $V\cdot T_{\text{p}}/T_{\text{e}}$ is calculated for both wavelengths. Fano parameters from Ref.~\cite{BDD_Pruvost} for a (001) MPCVD sample with a slightly lower doping level are shown together with present experimental results. The lower doping level in Ref.~\cite{BDD_Pruvost} is confirmed by the higher $\Omega_0$ resonant wavenumber, as the downshift is correlated with the boron incorporation \cite{BDD_Ushizawa}. (Note that the boron content is not equivalent to the carrier concentration, as boron dimers present in the material do not contribute to free carrier concentration \cite{BDD_Bustarret2008}.)\\
The negative sign of $V\cdot T_{\text{p}}/T_{\text{e}}$ is in agreement with Raman measurements in the visible range \cite{BDD_Pruvost,BDD_Wang2002}. The positive $V$ value was calculated \cite{BDD_Cardona}. The negative $T_{\text{p}}$ values are given in theoretical calculations \cite{BDD_Calleja}. The excitation energy, which is lower than the band-gap, contributes to positive $T_{\text{e}}$.\\
The Raman scattering cross section increases with the excitation frequency as $\sigma\propto \omega_L^4$, where $\omega_L$ is the incident photon frequency. The Raman tensor of the decoupled phonon is expected to follow $T_{\text{p}}\propto \omega_L^2$, while there is no similar relation for $T_{\text{e}}$. $V$ is independent of the photon excitation wavelength, so $\left|V\cdot T_{\text{p}}/T_{\text{e}}\right|$ might be expected to decrease with increasing incident wavelength. However, experimental data in Table \ref{Fano_summary} contradicts this expectation \cite{BDD_Pruvost}. In Table \ref{Fano_summary}, there is no typical wavelength dependence of the values. This anomaly, which is not supported by UV excitation measurements \cite{BDD_Locher,BDD_Ghodbane}, demands further investigation both in visible and UV range. In Ref.~\cite{BDD_Vlasov}, the anomalous wavelength dependence of $T_{\text{p}}$ was explained by the presence of regions with very high boron concentration, which result in breaking of the wavevector conservation rule and a strong enhancement in $T_{\text{p}}$ as approaching the plasmon frequency in the near infrared.\\
In Ref.~\cite{BDD_Pruvost}, $\left|T_{\text{p}}/V\cdot T_{\text{e}}\right|$ was found to be small. $V$ is large, as the deformation potential was obtained experimentally $d_0=$99.1 eV \cite{BDD_Calleja}. The present results in Table \ref{Fano_summary} show low values for $\left|V\cdot T_{\text{p}}/T_{\text{e}}\right|$ and yield an even stricter upper bound for $T_{\text{p}}$.
\section{Conclusions}
In summary, we performed Raman measurements in heavily boron doped nanocrystalline diamond. The Raman mode around 500 cm$^{-1}$ is assigned to point-defect boron dimers. The 1003 cm$^{-1}$ and the 1210 cm$^{-1}$ Raman modes are assigned to the phonon density of states. The 1210 cm$^{-1}$ Raman mode does not exhibit Fano lineshape.\\
The Fano line of the zone-center phonon of diamond is significantly red-shifted and indicates significant asymmetry ($q\approx-2$). The Raman scattering amplitude of the decoupled phonon is small and it shows an anomalous wavelength dependence.
\section{Acknowledgements}
Work supported by the ERC Grant Nr.\ ERC-259374-Sylo, and by the New Sz\'{e}chenyi Plan Nr.\ T\'{A}MOP-4.2.2.B-10/1.2010-0009.

\end{document}